\documentclass[aps,preprint,superscriptaddress]{revtex4}
\usepackage{graphicx}
\usepackage{amsmath}
\usepackage{dcolumn}
\usepackage{epstopdf}
\usepackage[caption=false]{subfig}

\newcommand{\bom}{\mbox{\boldmath $\bf\omega$}}

\newcommand{\bxi}{\mbox{\boldmath $\bf\xi$}}

\def\V{\textbf{V}}
\def\g{\textbf{g}}
\def\v{\textbf{v}}
\def\b{\textbf{b}}

\def\k{\textbf{k}}
\def\B{\textbf{B}}
\def\F{\textbf{F}}

\def\E{\textbf{E}}
\def\e{\textbf{e}}
\def\pa{\partial}

\def\r{\textbf{r}}

\begin{document}

\title{Modeling the Parker instability in a rotating plasma screw pinch}

\author{I. V. Khalzov}
\affiliation{University of Wisconsin, 1150 University Avenue, Madison, Wisconsin 53706 USA}
\author{B. P. Brown}
\affiliation{University of Wisconsin, 1150 University Avenue, Madison, Wisconsin 53706 USA}
\author{N. Katz}
\affiliation{University of Wisconsin, 1150 University Avenue, Madison, Wisconsin 53706 USA}
\author{C. B. Forest}
\affiliation{University of Wisconsin, 1150 University Avenue, Madison, Wisconsin 53706 USA}

\date{\today}

\begin{abstract}

We analytically and numerically study the analogue of the Parker (magnetic buoyancy) instability in a uniformly rotating plasma screw pinch confined in a cylinder. Uniform plasma rotation is imposed to create a centrifugal acceleration, which mimics the gravity required for the classical Parker instability. The goal of this study is to determine how the Parker instability could be unambiguously identified in a weakly magnetized, rapidly rotating screw pinch, in which the rotation provides an effective gravity and a radially varying azimuthal field is controlled to give conditions for which the plasma is magnetically buoyant to inward motion. We show that an axial magnetic field is also required to circumvent conventional current driven magnetohydrodynamic (MHD) instabilities such as the sausage and kink modes that would obscure the Parker instability.  These conditions can be realized in the Madison Plasma Couette Experiment (MPCX). Simulations are performed using the extended MHD code NIMROD for an isothermal compressible plasma model. Both linear and nonlinear regimes of the instability are studied, and the results obtained for the linear regime are compared with analytical results from a slab geometry. Based on this comparison, it is found that in a cylindrical pinch  the magnetic buoyancy  mechanism dominates at relatively large Mach numbers ($M>5$), while at low Mach numbers ($M<1$) the instability is due to the curvature of magnetic field lines. At intermediate values of Mach number ($1<M<5$) the Coriolis force has a strong stabilizing effect on the plasma. A possible scenario for experimental demonstration of the Parker instability in MPCX is discussed.       
\end{abstract}

\maketitle

\section{Introduction}

The Parker (or the magnetic buoyancy) instability arises in a stratified plasma, which is partially supported against gravity by a magnetic field. It was originally proposed by Parker that the magnetic buoyancy resulting in this system is a mechanism for the formation of sunspots  \cite{Parker_1955}. Since then the theory of the magnetic buoyancy instability has  been developed in numerous papers \cite{Newcomb_1961, Parker_1966, Yu_1966, Thomas_1975, Hughes_1985, Hughes_1987, Shibata_1990} and extended to include various effects, such as pressure anisotropy \cite{Yu_1967}, double diffusivity \cite{Schubert_1968, Roberts_1977, Acheson_1978a, Acheson_1979, Hughes_Weiss_1995},   rotation \cite{Roberts_1977, Acheson_1978a, Acheson_1979, Acheson_1978b, Zweibel_1975}, microturbulence \cite{Zweibel_1975}, line tying \cite{Zweibel_1992} and nonuniform gravity \cite{Kamaya_1997}. Full 3-D magnetohydrodynamic (MHD) simulations of the magnetic buoyancy instability have been performed in the solar context \cite{Abbett_et_al_2001, Fan_et_al_2003, Fan_2008, Vasil_Brummell_2008, Fan_2009, Weber_et_al_2011}, including some that explore the effects of double diffusivity \cite{Silvers_et_al_2009}. At present time this instability is accepted in the astrophysical community as the driver of  magnetic activity in objects ranging from stars to accretion disks and galaxies \cite{Tajima_2002}.  In the Sun, the Parker instability may even play a role in the global-scale magnetic dynamo \cite{Parker_1975, Cline_2003, Fan_2009, Charbonneau_2010}.    

Despite its important role in understanding the dynamics of the astrophysical magnetic fields, no direct experimental investigation of the Parker instability has been performed so far. The difficulty is in creating the proper conditions for this instability and, especially, in imitating relatively strong gravity in plasma.  Nonetheless, the Parker instability could be realized in experiments with rapidly rotating plasmas, where the role of effective gravity is played by centrifugal acceleration. This possibility of exciting the Parker instability  was pointed out in theoretical studies devoted to centrifugally confined plasmas \cite{Huang_2003, Huang_2004}. However, the instability was not observed in the corresponding experiment.  The reason lies, perhaps, in the profiles of magnetic field and rotation, which are optimized for plasma confinement and, therefore, for suppressing the instabilities.  Thus, the experimental observation of the Parker instability is still lacking.  

\begin{figure}[tbp]
\begin{center}
  \includegraphics[scale=0.8]{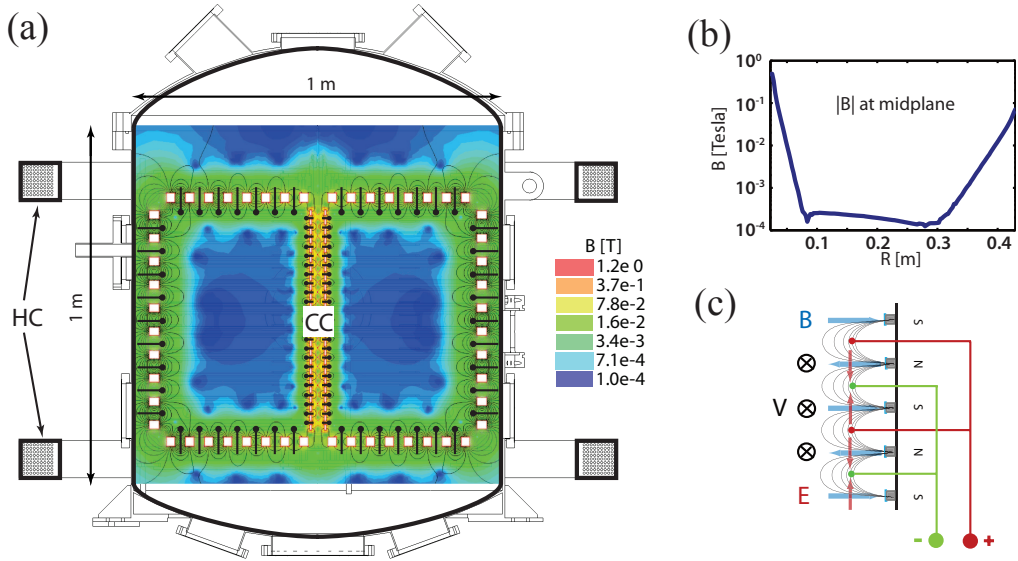}\\
  \caption{Madison Plasma Couette Experiment (MPCX): (a) sketch and magnetic field amplitude; (b) typical radial profile of cusp magnetic field; (c) electrode configuration near wall. Rings of permanent magnets of alternating polarity line the inside of the cylinder with their poles oriented normally to the walls. Electrodes are placed between the magnets. The azimuthal velocity at the boundary can be adjusted through variation of the $\textbf{E}\times\textbf{B}$ drift by changing the voltages between electrodes. Helmholtz coils (HC) are used to induce external (mainly axial) magnetic field. Removable center core (CC) is shown, but it will not be used in the Parker experiment. Figure is adopted from Ref. \cite{Fatima_2011}}\label{MPCX}
\end{center}
\end{figure}

Our present paper is motivated by the recent results from the Madison Plasma Couette Experiment (MPCX)  showing a controllable plasma rotation \cite{APS_2011}. The MPCX has been specifically designed  to study phenomena associated with plasma flows \cite{APS_2008}. A unique experimental setup is implemented in MPCX: a multicusp magnetic field localized near the walls of the cylindrical vessel provides plasma confinement and along with the applied electric field from the wall electrodes drives the prescribed azimuthal flows (Fig.~\ref{MPCX}). Different types of flows potentially achievable in the MPCX can be used for investigation of such phenomena as magnetic dynamo \cite{Ivan_2011}  and magnetorotational instability \cite{Fatima_2011}.
 
The goal of this theoretical study is to determine the plasma parameters, flows and magnetic fields  required for excitation of the Parker instability in a cylindrical geometry and to demonstrate the possibility of obtaining such instability in the MPCX.  Consideration is performed in the frame-work of isothermal compressible magnetohydrodynamics  (MHD), which appears to be  a good approximation for the MPCX plasma. The numerical results reported in the paper are obtained using the extended MHD code NIMROD \cite{Sovinec_2004}, which can accurately model both linear and nonlinear plasma dynamics in  a specific geometry for realistic experimental conditions.  The NIMROD code has been used previously to study dynamo action  \cite{Ivan_2011} and magnetorotational instability \cite{Fatima_2011} in MPCX.

In the paper we follow the convention of Ref. \cite{Tajima_2002} and use the term ``Parker instability" to denote an \textit{undular mode} of magnetic buoyancy instability, in which the wave vector has a component parallel to the magnetic field and, therefore, the perturbed magnetic field lines are bent. This mode is opposed  to an \textit{interchange mode}, in which the wave vector is perpendicular to the magnetic field and perturbations do not bend the field lines. It turns out that in a plasma screw pinch with rigid-body rotation the stability boundaries are determined by the modes with non-zero parallel component of the wave vector. These field-bending modes are of primary interest in the paper. 

The detailed analysis of the Parker instability was performed by Newcomb \cite{Newcomb_1961}. He considered a perfectly conducting plasma in a varying with height horizontal magnetic field and a uniform  vertical gravitational field [similar to configuration shown in Fig. \ref{geom}(a)  for $X>0$, with gravity acting in the positive $X$ direction $\textbf{G}=G_0\e_x$, and magnetic field having the only component  $B_y(X)$]. The equilibrium force balance is 
\begin{equation}
\label{new_eq}
\frac{d}{dX}\bigg(P+\frac{B_y^2}{8\pi} \bigg)=\rho G_0,
\end{equation}
where $P$ is the plasma pressure and $\rho$ is the plasma density. According to Newcomb, the system is unstable with respect to the Parker instability if  for some perturbation $\xi_x$ the energy integral is negative,
\begin{equation}
\label{new_stab}
W=\frac{1}{2}\int\bigg( \frac{K_y^2B_y^2}{4\pi(K_y^2+K_z^2)}\,\bigg|\frac{\pa \xi_x}{\pa X}\bigg|^2+\frac{K_y^2B_y^2}{4\pi}|\xi_x|^2 +\bigg( G_0\frac{d\rho}{dX}-\frac{\rho^2G_0^2}{\gamma P}\bigg)\, |\xi_x|^2\bigg)\,d^3\r<0.
\end{equation}
Here $\gamma$ is the adiabatic index, $K_y$ and $K_z$ are components of the wave vector of the perturbation.  The first two terms in the integral are always positive and  represent  the stabilizing effect of magnetic field lines bending. The instability can arise only when the third term is negative. In an isothermal plasma with adiabatic index $\gamma=1$ and $P=\rho C_{s}^2$, where $C_{s}$ is the average sound speed,  Eq. (\ref{new_stab}) is reduced to
\begin{equation}
\label{new_stab2}
W=\frac{1}{8\pi}\int\bigg( \frac{K_y^2B_y^2}{K_y^2+K_z^2}\,\bigg|\frac{\pa \xi_x}{\pa X}\bigg|^2+K_y^2B_y^2|\xi_x|^2 -\frac{G_0}{2 C_{s}^2}\frac{dB_y^2}{dX}\, |\xi_x|^2\bigg)\,d^3\r<0,
\end{equation}
which means that the magnetic pressure must increase in the direction of gravity for the system to be unstable. A simple physical interpretation of this is given by Acheson \cite{Acheson_1979}. In equilibrium, the magnetic pressure supports more mass against gravity than it would be possible in its absence. This extra mass, expressed in the form of magnetic pressure,  enters the destabilizing third term in Eq. (\ref{new_stab2}).  The situation is somewhat analogous to the Rayleigh-Taylor instability, in which the heavy fluid lies on top of a lighter one \cite{Chandra_1961}.

As seen from Eq. (\ref{new_stab2}), two key elements are necessary for driving the magnetic buoyancy instability in plasma: gravity and magnetic field. Below we outline some specific conditions, which these elements should satisfy in the experiment.

\begin{enumerate}
\item[(1)] \textit{Gravity (rotation)}. Centrifugal acceleration arising in a rotating plasma is the efficient way to mimic gravity in the experiment. Obvious disadvantages of such an approach are the possible excitation of unwanted instabilities related to the rotation profile, such as Rayleigh instability, Kelvin-Helmholtz instability, magnetorotational instability (MRI). A further consequence of rotation is that we might expect the Coriolis force to stabilize some motions. If we assume that the rotation is uniform in the axial direction, then  the most limiting condition on radial profile of angular velocity $\Omega(R)$ comes from the MRI: to have a stable rotation in the presence of axial magnetic field one needs to satisfy $\pa(\Omega^2)/\pa R\geq0$ \cite{Velikhov_1959, Chandra_1960}. We consider here only a case of rigid-body rotation with $\Omega(R)=\Omega_0=\textrm{const}$ corresponding to effective centrifugal gravity $\textbf{G}=\Omega_0^2R\,\e_r$.  Such plasma rotation has been successfully obtained in recent MPCX runs \cite{APS_2011}.     
\item[(2)] \textit{Magnetic field}. Only internal (i.e., induced by currents flowing in plasma) magnetic fields can be potentially buoyant. This is because magnetic fields contribute to the plasma force balance only if the corresponding Lorentz force is not zero. Evidently, externally applied fields do not have this property, so they alone cannot induce Parker instability. In addition, the magnitude of the field (magnetic pressure) has to be increasing in the direction of gravity. One of the feasible  ways to create such magnetic field in  the MPCX is to drive axial current through the plasma, so that azimuthal field is  induced. However, the curvature of a purely azimuthal magnetic field makes plasma susceptible to the so-called ``sausage" instability (with azimuthal mode number $m=0$) \cite{Bateman_1978}. This is obviously unwanted for the Parker instability experiment, as this instability may mask the buoyancy instability. To avoid this complication, one can add an external axial magnetic field (by driving current in the Helmholtz coils, see Fig.~\ref{MPCX}); this leads to a screw pinch with a helical magnetic field. Such magnetic field is also unstable to parasitic kink instabilities \cite{Bateman_1978}, but as we will show, the regions of plasma parameters corresponding to the kink  modes and the Parker instability are well separated, and these two instabilities can be easily distinguished in the experiment.  In the following we consider only a case with uniform axial current density (which is consistent with assumptions of uniform resistivity profile in isothermal plasma) and uniform axial magnetic field, so the radial profiles of field components are $B_\varphi(R)=B_aR/a$ and $B_z(R)=B_{z0}=\textrm{const}$, where  $B_a$ and $B_{z0}$ are constant and $a$ is the radius of the pinch (Fig. \ref{geom}). Thus, the main object of our study is the plasma screw pinch with constant pitch and rigid-body rotation.  
\end{enumerate}

We note here that unfavorable curvature of the helical field resulting from cylindrical geometry of the screw pinch can lead to development of unwanted kink instabilities. To tell whether the instability is due to kink modes or  magnetic buoyancy, we also perform the stability analysis of analogous plasma configuration  in a slab geometry, where the effects associated with the field curvature are absent and only magnetic buoyancy can play a role. Then we compare and contrast the results obtained in the two geometries. 

The structure of the paper is as follows. In Sec. II, we describe the model used in our study. In Sec. III, the Parker instability is investigated in a slab geometry and the stability boundaries are obtained. In Sec. IV, the linear stability analysis of a plasma screw pinch with rigid-body rotation is performed in a geometry of periodic cylinder and the region of parameters appropriate for the Parker instability is determined. In Sec. V, we study the more realistic case of a bounded cylinder with inclusion of dissipative effects. With NIMROD simulations, we explore the nonlinear dynamics of the Parker instability as well. In Sec. VI, we summarize and discuss the possible scenarios for experimental demonstration of the Parker instability in the MPCX.

\section{Model}

\begin{figure}[tbp]
\begin{center}
  \includegraphics[scale=1]{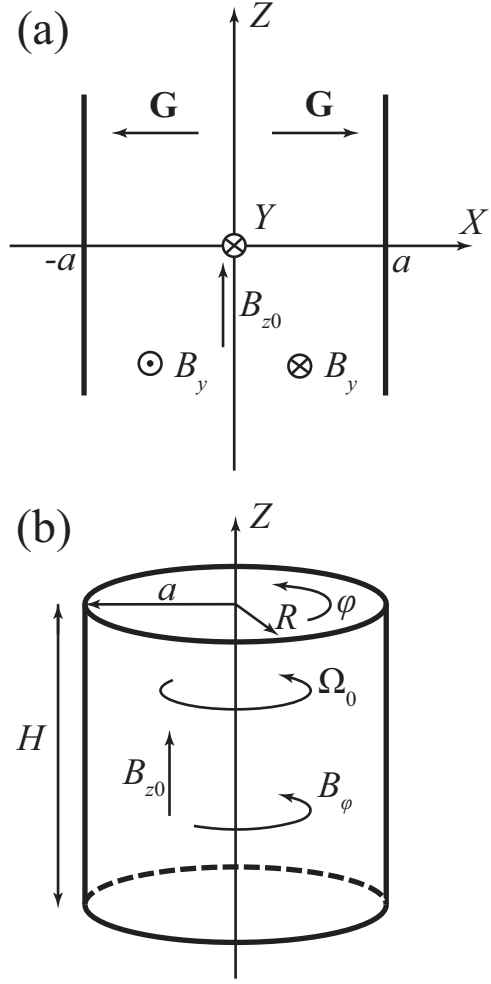}\\
  \caption{Geometries of the problem: (a) slab; (b) cylinder.}\label{geom}
\end{center}
\end{figure}

We consider the stability problem for two equivalent plasma configurations in two different geometries -- slab and cylinder (Fig. \ref{geom}). This is done to distinguish  the Parker instability in the cylinder from kink instabilities related to the field curvature, as in our slab geometry there is no curvature in the imposed magnetic field. The corresponding equilibrium fields and gravity are given by
\begin{eqnarray}
\label{eq_slab}&&\textrm{Slab}:~~~~~~~~~ \textbf{V}_{eq}=0,~~~~~~~~~~\textbf{B}_{eq}=B_a\,\frac{X}{a}\e_y+B_{z0}\e_z,~~~\textbf{G}=\Omega_0^2\,X\,\e_x;\\
&&\label{eq_cyl}\textrm{Cylinder}:~~~ \textbf{V}_{eq}=\Omega_0\,R\e_\varphi,~~~\textbf{B}_{eq}=B_a\,\frac{R}{a}\e_\varphi+B_{z0}\e_z,~~~\textbf{G}=0.
\end{eqnarray} 
Note that the gravity in slab geometry is chosen in such way to be analogous to the centrifugal acceleration in cylinder. In both cases the background plasma is generally stratified; the exact forms of the density profiles are determined by the force balance equation and will be given in Secs. III and IV for the respective cases.       

As a frame-work for our study we use isothermal compressible MHD model, which in nondimensional form is
\begin{eqnarray}
\label{mhd1}
\frac{\pa n}{\pa\tau}&=&-\nabla\cdot(n\v),\\
\label{mhd2}
n\frac{\pa \v}{\pa\tau}&=&-n(\v\cdot\nabla)\v - \frac{\beta}{2}\nabla n+(\nabla\times\b)\times\b+n\g + \nu\bigg(\nabla^2\v+\frac{1}{3}\nabla(\nabla\cdot\v)\bigg),\\
\label{mhd3}
\frac{\pa\b}{\pa\tau}&=&\nabla\times(\v\times\b)+\eta\nabla^2\b.
\end{eqnarray}
As the motions are themselves isothermal, no energy equation is solved.  This corresponds to the limit where thermal conduction is much faster than plasma motions, which is the appropriate limit for MPCX.  Magnetically buoyant motions in this system correspond to the limit of doubly-diffusive motions \cite{Acheson_1979}, and as such, these isothermal motions are likely to be maximally unstable to magnetic buoyancy. In Eqs. (\ref{mhd1})-(\ref{mhd3}) the unit of length is the characteristic size of the system $a$ (half-width of the slab layer, radius of the cylinder), $\tau$, $n$, $\v$, $\b$ and $\g$ stand for normalized time, density, velocity, magnetic field and gravity, respectively: 
\begin{equation}
\label{norm}
\tau=\frac{V_A}{a}\,t,~~~n=\frac{\rho}{\rho_0},~~~\v=\frac{\V}{V_A},~~~\b=\frac{\B}{B_{z0}},~~~\g=\frac{a\textbf{G}}{V_A^2},
\end{equation}
where $V_A=B_{0z}/\sqrt{4\pi\rho_0}$ is the  Alfven velocity based on the externally applied magnetic field $B_{z0}$ (vertical in slab, axial in cylinder), $\rho_0$ is the average mass density. The thermal to magnetic pressure ratio $\beta$, dimensionless viscosity $\nu$ and resistivity $\eta$ are defined as 
$$
\beta=\frac{8\pi P_0}{B_{z0}^2},~~~\nu=\frac{\mu}{\rho_0 aV_A},~~~\eta=\frac{c^2}{4\pi\sigma aV_A},
$$
where $P_0$ is the average thermal pressure, $\mu$ is the dynamic viscosity, $\sigma$ is the plasma conductivity and $c$ is the speed of light;  these parameters are  assumed to be constant in time and uniform in space. For complete description of equilibrium configurations we also introduce Mach number $M$ and pinch parameter $\theta$:
$$
M=\frac{V_0}{C_s},~~~\theta=\frac{B_a}{B_{z0}},
$$
where $V_0=a\Omega_0$ is the velocity at the cylindrical boundary (peak driving velocity), $C_s=\sqrt{P_0/\rho_0}$ is the average plasma  sound speed with adiabatic index $\gamma=1$ for isothermal model and $B_a$ is  the boundary value of the magnetic field component, which is induced by vertical (axial) current in plasma.

The MPCX plasma parameters can be varied in a wide range allowing experimentalists a great flexibility in choosing the regimes of operation (Table \ref{t1}). The dependencies  of the non-dimensional quantities used in our study on the plasma parameters are presented below: thermal to magnetic pressure ratio
\begin{equation}
\label{beta}
\beta\equiv\frac{8\pi P_0}{B_{z0}^2}=40\,\frac{N_0[10^{18}~\textrm{m}^{-3}]\,(T_e[\textrm{eV}]+T_i[\textrm{eV}])}{B_{z0}^2[\textrm{G}]},
\end{equation}
normalized viscosity (inverse fluid Reynolds number based on Alfven velocity)
\begin{equation}
\label{visc}
\nu\equiv\frac{\mu}{\rho_0aV_A}=0.88\,\frac{T^{5/2}_i[\textrm{eV}]} {a[\textrm{m}]\,B_{z0}[\textrm{G}]\,N_0^{1/2}[10^{18}~\textrm{m}^{-3}]\,\lambda},
\end{equation}
normalized resistivity (inverse Lundquist number)
\begin{equation}
\label{res}
\eta\equiv\frac{c^2}{4\pi\sigma aV_A}=0.019\,\frac{N_0^{1/2}[10^{18}~\textrm{m}^{-3}]\,\mu_i^{1/2}\,\lambda} {a[\textrm{m}]\,B_{z0}[\textrm{G}]\,T_e^{3/2}[\textrm{eV}]},
\end{equation}
Mach number 
\begin{equation}
\label{mach}
M\equiv\frac{V_0}{C_s}=0.10\,\frac{V_0[\textrm{km/s}]\,\mu_i^{1/2}} {(T_e[\textrm{eV}]+T_i[\textrm{eV}])^{1/2}},
\end{equation} 
where $\lambda$ is the Coulomb logarithm (typically $\lambda\approx10-20$). Eqs. (\ref{visc}) and (\ref{res}) are derived from the Braginskii equations for a plasma with singly charged ions in a weak magnetic field \cite{Brag_1965}. The typical values of these quantities in the MPCX are also presented in Table \ref{t1}.  

\begin{table}[tbp]
\caption{Parameters of MPCX}
\begin{center}
\begin{tabular}{lccc}
\hline
\hline
Quantity & Symbol & Value & Unit \\
\hline 
Radius of cylinder & $a$ & 0.5 & m \\
Height of cylinder & $H$ & 1 & m \\
Peak driving velocity  & $V_0$ & $0-20$ & km/s\\
Axial magnetic field  & $B_{z0}$ & $0-100$ & G\\
Average number density &   $N_0$  &  $10^{16}-10^{17}$ & m$^{-3}$\\
Electron temperature & $T_e$ & $2-10$ & eV \\
Ion temperature & $T_i$ &  $0.2-2$ & eV \\
Ion species &  & H, He, Ne, Ar &\\
Ion mass & $\mu_i$ & 1, 4, 20, 40 & amu\\
\hline
Thermal/magnetic pressure ratio  & $\beta$ & $\geq8.8\times10^{-5}$ & \\
Normalized viscosity & $\nu$ & $\geq2.1\times10^{-4}$ & \\
Normalized resistivity & $\eta$ & $\geq1.2\times10^{-5}$ & \\
Mach number & $M$ & $0-8.5$ & \\
\hline
\end{tabular}
\end{center}
\label{t1}
\end{table}

Note that the cylindrical equilibrium configuration in Eqs. (\ref{eq_cyl}) contains neither the multicusp magnetic field nor the boundary layers present in the real experiment. In our study we do not focus on the details of the plasma driving and simply assume that the equilibrium rigid-body plasma rotation is given a priori. This assumption is consistent with recent observations showing such rotation in the bulk of the MPCX plasma \cite{APS_2011}. 

Eqs. (\ref{mhd1})-(\ref{mhd3}) must be supplemented with boundary conditions. Here we assume impenetrable, perfectly conducting  walls, so the normal components of the velocity and the time-varying magnetic field vanish at the boundary $\Gamma$:
\begin{equation}
\label{bc_ideal}
v_n|_{\Gamma}=0,~~~\tilde{b}_n|_{\Gamma}=0.
\end{equation}
These conditions are enough in the ideal MHD cases (when $\eta=\nu=0$) considered in Secs. III and IV. In the dissipative MHD case (when $\eta\neq0$ and $\nu\neq0$) considered in Sec. V, two additional conditions are required, namely, no-slip condition for the tangential velocity and absence of the tangential time-varying electric field:   
\begin{equation}
\label{bc_nonideal}
\v_t|_{\Gamma}=(\v_{eq})_t|_{\Gamma},~~~\tilde{\E}_t|_{\Gamma}=\eta(\nabla\times\tilde\b)_t|_{\Gamma}=0,
\end{equation}
where $\v_{eq}$ is the equilibrium velocity, and tangential electric field at the boundary is determined only by the corresponding component  of the current. In the following sections we study the stability of the equilibrium configurations [Eqs. (\ref{eq_slab}), (\ref{eq_cyl})] in the frame-work of the isothermal MHD model [Eqs. (\ref{mhd1})-(\ref{mhd3})] with appropriate boundary conditions given by Eqs. (\ref{bc_ideal}), (\ref{bc_nonideal}).

\section{Slab: ideal MHD stability}

\begin{figure}[tbp]
\begin{center}
  \includegraphics[scale=1]{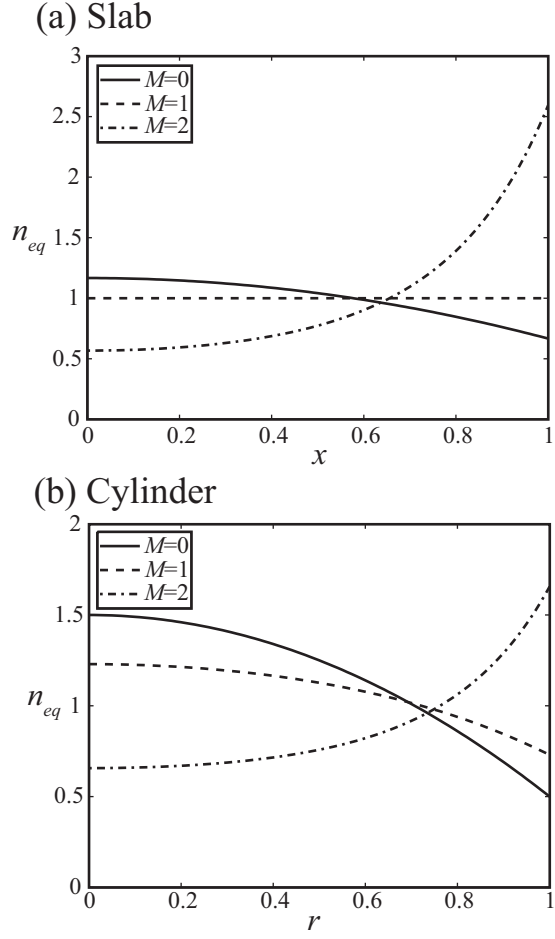}\\
  \caption{Equilibrium density profiles for $\beta=8$, pinch parameter $\theta=2$ and different values of Mach number $M$: (a) in slab [Eq. (\ref{slab_n})]; (b) in cylinder [Eq. (\ref{cyl_n})].}\label{den}
\end{center}
\end{figure}

First, we consider the ideal MHD stability (no dissipation, $\nu=\eta=0$) of a stratified magnetized plasma in a slab geometry as shown in Fig. \ref{geom}(a).  In this periodic slab geometry, there are no Coriolis forces and curvature effects associated with the imposed magnetic field that will be present in cylindrical geometries, so the magnetic buoyancy is the only possible destabilizing mechanism.  The normalized equilibrium fields and gravity in this case are 
\begin{equation}
\label{slab_eq}\v_{eq}=0,~~~\b_{eq}=\theta x\,\e_y+\e_z,~~~\g=g(x)\,\e_x=\frac{\beta}{2}\,M^2x\,\e_x.
\end{equation} 
The layer is bounded in the $x$ direction ($-1<x<1$), and periodic in $y$ and $z$. The $x$ component of the force balance results in an equation for the equilibrium density profile $n_{eq}(x)$:
\begin{equation}
\label{slab_fb}\frac{\beta}{2}\,\frac{dn_{eq}}{dx}=\frac{\beta}{2}\,M^2xn_{eq}-\theta^2x.
\end{equation} 
Taking into account that the average normalized density is 1, we find the solution to Eq. (\ref{slab_fb}) in the form:
\begin{equation}
\label{slab_n} n_{eq}(x)=\bigg(1-\frac{2\theta^2}{\beta M^2}\bigg) \frac{\sqrt{2}M\,e^{M^2x^2/2}}{\sqrt{\pi}\,\textrm{erfi}(M/\sqrt{2})}+\frac{2\theta^2}{\beta M^2},
\end{equation} 
where \textit{erfi} is the so-called imaginary error function defined via standard error function \textit{erf} as 
$$
\textrm{erfi}\,(z)\equiv-i\,\textrm{erf}\,(i\,z)=\frac{2}{\sqrt{\pi}}\int\limits_0^z\,e^{t^2}\,dt.
$$
The density profiles corresponding to Eq. (\ref{slab_n}) are shown in Fig. \ref{den}(a) for several values of Mach number.

We linearize Eqs. (\ref{mhd1}-\ref{mhd3}) near the equilibrium given by Eqs. (\ref{slab_eq}), (\ref{slab_n}) and introduce the plasma displacement vector $\bxi(x,~y,~z) e^{-i\omega\tau}$ to describe the perturbations of the physical quantities:
$$
\delta n=-\nabla\cdot(n_{eq}\bxi),~~~\delta\v=-i\omega\bxi,~~~\delta\b=\nabla\times(\bxi\times\b_{eq}).
$$
The linearized equation for the displacement $\bxi$ is  
\begin{equation}
\label{slab_xi} -\omega^2 n_{eq}\bxi=\F[\bxi]\equiv-\frac{\beta}{2}\nabla\delta n+(\nabla\times\delta\b)\times\b_{eq}+(\nabla\times\b_{eq})\times\delta\b +\delta n\,\g.
\end{equation}
Due to periodicity of the slab in $y$ and $z$, we can assume $\bxi$ depends on $y$ and $z$ as $e^{ik_yy+ik_zz}$ and consider stability of modes with different wave numbers $(k_y,~k_z)$ separately. The boundary conditions on $\bxi$ in the $x$ direction follow from  Eq. (\ref{bc_ideal}):
\begin{equation}
\label{slab_BC} \xi_x|_{x=\pm1}=0.
\end{equation}
Eqs. (\ref{slab_xi}) and (\ref{slab_BC}) constitute an eigenvalue problem.  

In this relatively simple system, without solving the full eigenvalue problem one can obtain the stability criterion analytically using  the energy principle \cite{BFKK} . It states that the static (without flows) ideal MHD system is linearly stable if and only if the potential energy of the perturbation, 
\begin{equation}
\label{W} W=-\frac{1}{2}\int\bxi^*\cdot\F[\bxi]\,d^3\r,
\end{equation}
is positive for all displacements $\bxi$ satisfying the boundary conditions in the problem (provided that the operator $\F[\bxi]$ is self-adjoint). Here displacements are assumed to be complex in general and the star denotes the complex conjugation.   

A general expression for the potential energy of the perturbation with wave number $\k=(k_y,~k_z)$ in a slab is
\begin{eqnarray}
\label{slab_W}
W&=&\frac{1}{2}\int\bigg(\bigg(\frac{\beta}{2}\,n_{eq}+\b_{eq}^2\bigg)|\nabla\cdot\bxi|^2 + F^2|\bxi|^2 + iF\big((\b_{eq}\cdot\bxi^*)(\nabla\cdot\bxi)-(\b_{eq}\cdot\bxi)(\nabla\cdot\bxi^*)\big)\nonumber\\
&+&gn_{eq}\big(\xi_x(\nabla\cdot\bxi^*)+\xi^*_x(\nabla\cdot\bxi)\big)+g\frac{dn_{eq}}{dx}|\xi_x|^2 \bigg)\,d^3\r,
\end{eqnarray}
where $F=\k\cdot\b_{eq}=k_z+k_y\theta x$ characterizes the component of the wave vector parallel to the equilibrium field. The functional given by Eq. (\ref{slab_W}) is self adjoint, so the energy principle applies.    
  
Assuming that $F$ is not identically equal to zero on the interval $-1<x<1$, we minimize $W$ with respect to $\xi_y$ and $\xi_z$. The system is unstable if for some $\xi_x$ the minimized potential energy is negative, i.e., 
\begin{equation}
\label{slab_Wmin}
W_{min}=\frac{1}{2}\int\bigg( \frac{F^2}{k_y^2+k_z^2}\,\bigg|\frac{\pa \xi_x}{\pa x}\bigg|^2+F^2|\xi_x|^2-M^2\theta^2x^2|\xi_x|^2\bigg)\,d^3\r<0,
\end{equation}
in agreement with Ref. \cite{Newcomb_1961} and Eq. (\ref{new_stab2}). Note that the only destabilizing term in Eq. (\ref{slab_Wmin}) is due to joint effect of gravity and plasma current (or magnetic pressure gradient); this instability condition does not depend on the density profile explicitly. For modes with $k_y=0$ (which are the analogue of axisymmetric modes with $m=0$ in a cylinder) the instability condition is simplified to 
\begin{equation}
\label{slab_MS} \theta^2M^2>k_x^2+k_z^2,
\end{equation}
where $k_x$ is the effective wave number in $x$ direction. According to Eq. (\ref{slab_MS}), the system becomes unstable when magnetic buoyancy (the term on the left-hand side) overcomes magnetic tension (the term on the right-hand side).

\begin{figure}[tbp]
\begin{center}
  \includegraphics[scale=0.9]{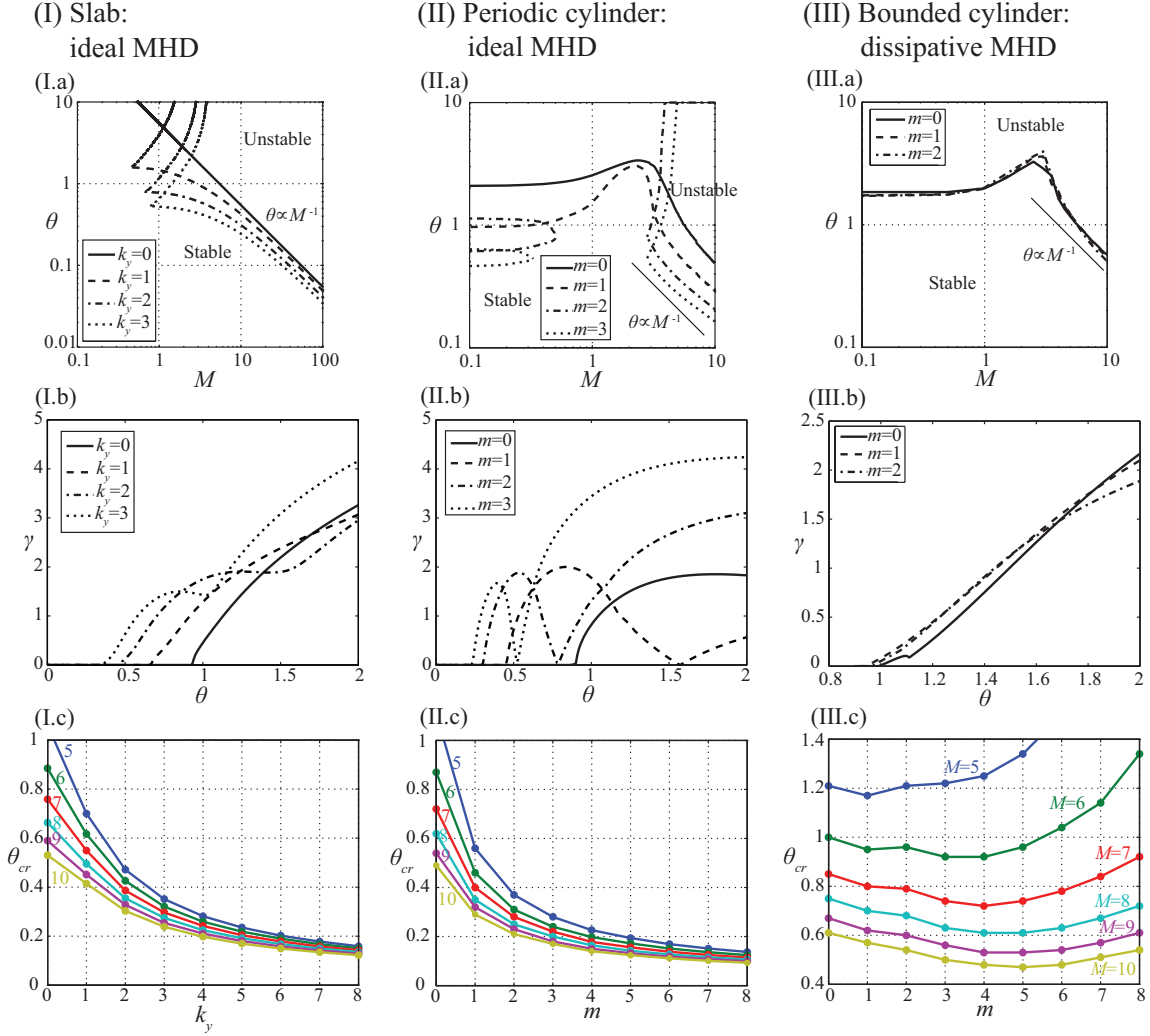}\\
  \caption{Results of numerical solution to eigenvalue problems for $\beta=8$ in (I) slab with $k_z=-\pi/2$, (II) periodic cylinder with $k_z=-\pi/2$ and (III) bounded cylinder with viscosity $\nu=0.01$ and resistivity $\eta=0.01$: (a) marginal stability curves on the plane of Mach number--pinch parameter ($M-\theta$); (b) growth rate of instability $\gamma=\textrm{Im}~\omega$ as function of pinch parameter $\theta$ for Mach number $M=6$; (c) critical pinch parameter $\theta_{cr}$ as function of $k_y$ (in slab) or $m$ (in cylinder) for different Mach numbers $M$ (indicated next to the curves).}\label{main}
\end{center}
\end{figure} 

This analytical consideration is confirmed by the results of the numerical solution to eigenvalue problem  [Eqs. (\ref{slab_xi}), (\ref{slab_BC})] presented in Fig.  \ref{main}(I.a-c). As one can see from Fig.  \ref{main}(I.a), the behavior of the marginal stability curves are similar for all values of $k_y$  in the region of large Mach numbers $M>5$. In that region, the scaling for the critical (required for the instability) value of the pinch parameter is 
\begin{equation}
\label{Q_cr}
\theta_{cr}\propto M^{-1}, 
\end{equation}
which is in agreement with Eq. (\ref{slab_MS}). Thus, we can conclude that such scaling is a universal signature of the Parker instability  in a slab geometry with equilibrium given by Eqs. (\ref{slab_eq}), (\ref{slab_n}).  
As follows from Fig. \ref{main}(I.b), when $\theta>\theta_{cr}$ the instability develops on the Alfven time scale $t_A=a/V_A$ [growth rate $\gamma=\textrm{Im}~\omega$ is normalized according to Eq. (\ref{norm})]. From Fig. \ref{main}(I.c) we see that  the critical pinch parameter $\theta_{cr}$ decreases for higher wave numbers $k_y$ (shorter wave lengths). This result appears only in ideal MHD consideration; as we show below, in dissipative MHD system the modes with shorter wave lengths are stabilized.      

\section{Periodic cylinder: ideal MHD stability}

In this section we study an ideal MHD stability (no dissipation, $\nu=\eta=0$) of a rotating plasma in a cylindrical screw pinch geometry shown in Fig. \ref{geom}(b). In this geometry of a periodic cylinder, the Coriolis forces and curvature forces associated with the imposed magnetic field are self-consistently included. The normalized equilibrium fields and gravity  in this case are 
\begin{equation}
\label{cyl_eq}\v_{eq}=\omega_0r\e_\varphi=\sqrt{\frac{\beta}{2}}\,Mr\,\e_\varphi,~~~\b_{eq}=\theta r\,\e_\varphi+\e_z,~~~\g=0.
\end{equation}
Plasma is bounded by a rigid, perfectly conducting cylindrical wall at $r=1$. To simplify the analysis we assume that the cylinder is periodic in  $z$ (axial) directions. The equilibrium density is determined from the force balance equation:
\begin{equation}
\label{cyl_fb}\frac{\beta}{2}\,\frac{dn_{eq}}{dr}=\frac{\beta}{2}\,M^2r\,n_{eq}-2\theta^2r.
\end{equation} 
The solution is
\begin{equation}
\label{cyl_n} n_{eq}(r)=\frac{M^2}{2}\bigg(1-\frac{4\theta^2}{\beta M^2}\bigg) \frac{e^{M^2r^2/2}}{e^{M^2/2}-1}+\frac{4\theta^2}{\beta M^2},
\end{equation} 
where we have taken into account that the average normalized density is 1. Samples of the density profiles given by Eq. (\ref{cyl_n}) are shown in Fig. \ref{den}(b).

We linearize Eqs. (\ref{mhd1}-\ref{mhd3}) near the equilibrium given by Eqs. (\ref{cyl_eq}), (\ref{cyl_n}) and consider all perturbations in a reference frame rotating with equilibrium angular velocity $\omega_0=\sqrt{\beta/2}\,M$. Since the system is periodic in azimuthal and axial directions we take the dependences of perturbations on $\varphi$  and $z$ as $e^{im\varphi+ik_zz}$ and study the stability of modes with different $(m,~k_z)$ separately. The transition to a rotating reference frame mathematically corresponds to changing the eigenfrequency of a mode with given $m$  according to the rule: $\bar\omega=\omega-im\omega_0$, where $\bar\omega$ and $\omega$ are the eigenfrequencies in the rotating and non-rotating reference frames, respectively. Introducing the plasma displacement vector $\bxi(r,~\varphi,~z)e^{-i\bar\omega\tau}$ we then obtain
$$
\delta n=-\nabla\cdot(n_{eq}\bxi),~~~\delta\v=-i\bar\omega\bxi,~~~\delta\b=\nabla\times(\bxi\times\b_{eq}),
$$
i.e., the expressions for perturbations of the physical quantities in terms of $\bxi$ are exactly the same as in a static case. In the rotating reference frame, the linearized equation for the displacement vector $\bxi$ is now: 
\begin{equation}
\label{cyl_xi} -\bar\omega^2n_{eq}\bxi+2n_{eq}\bom_0\times\delta\v =\F[\bxi]\equiv-\frac{\beta}{2}\nabla\delta n+(\nabla\times\delta\b)\times\b_0+(\nabla\times\b_0)\times\delta\b +\delta n\,\omega_0^2\,r\,\e_r,
\end{equation}
where $\bom_0=\omega_0\e_z$. The boundary conditions on $\bxi$ follow from  Eq. (\ref{bc_ideal}):
\begin{equation}
\label{cyl_BC} \xi_r|_{r=1}=0.
\end{equation}
Comparing Eqs. (\ref{slab_xi}) and (\ref{cyl_xi}), we see that in a system with plasma rotation there is a new effect due to the Coriolis force (the term with $\delta\v$), while the  centrifugal acceleration plays the role of gravity.

Due to the presence of the Coriolis force  in Eq. (\ref{cyl_xi}) the energy principle does not give a stability criterion in this case. However, as shown in Ref. \cite{Frieman_1960} it still can be applied to obtain a sufficient stability condition. Neglecting the Coriolis term, we write the potential energy of the perturbation with wave vector $\k=(m/r,~k_z)$ as 
\begin{eqnarray}
\label{cyl_W}
W&=&\frac{1}{2}\int\bigg(\bigg(\frac{\beta}{2}\,n_{eq}+\b_{eq}^2\bigg)|\nabla\cdot\bxi|^2 + F^2|\bxi|^2 + iF\big((\b_{eq}\cdot\bxi^*)(\nabla\cdot\bxi)-(\b_{eq}\cdot\bxi)(\nabla\cdot\bxi^*)\big)\nonumber\\
&+&\frac{\beta}{2}\frac{dn_{eq}}{dr}\big(\xi_r(\nabla\cdot\bxi^*)+\xi^*_r(\nabla\cdot\bxi)\big)+\omega_0^2\,r\frac{dn_{eq}}{dr}|\xi_r|^2+2i\theta F\big(\xi_\varphi\xi_r^*-\xi_\varphi^*\xi_r\big) \bigg)\,d^3\r,
\end{eqnarray}
where $F=\k\cdot\b_{eq}=k_z+m\theta$. Assuming that $F\neq0$, we minimize $W$ with respect to $\xi_\varphi$ and $\xi_z$. The result is that the system can become unstable if for some $\xi_r$ 
\begin{equation}
\label{cyl_Wmin}
W_{min}=\frac{1}{2}\int\bigg( \frac{F^2}{k^2r^2}\,\bigg|\frac{\pa(r\xi_r)}{\pa r}\bigg|^2+F^2|\xi_r|^2 - 2M^2\theta^2r^2|\xi_r|^2-\frac{4k_z^2\theta(\theta k^2r^2+mF)}{k^4r^2}\,|\xi_r|^2\bigg)\,d^3\r<0,
\end{equation}
where $k^2=m^2/r^2+k_z^2$. Note that there are two possible destabilizing mechanisms now: one of them is the same as in the slab geometry, and is due to the joint effect of rotation and plasma current. This is the Parker instability.  The second instability, absent in the slab geometry, is solely due to the plasma current, and can exist in the system without rotation.  These are ``sausage" or kink instabilities, depending on the azimuthal wave number $m$. For axisymmetric modes with $m=0$, Eq. (\ref{cyl_Wmin}) yields the instability condition in the following form:         
\begin{equation}
\label{cyl_MS} 2\theta^2(M^2+2)>k_r^2+k_z^2,
\end{equation}
where  $k_r$ is some effective radial wave number. One can see from Eq. (\ref{cyl_MS}), that at large Mach numbers $M\gg1$ the critical value of the pinch parameter scales as $\theta_{cr}\propto1/M$, which is identical to Eq. (\ref{Q_cr}). This suggests that the Parker destabilization mechanism dominates in the cylindrical pinch when $M\gg1$ if the Coriolis force is ignored. 

\begin{figure}[tbp]
\begin{center}
  \includegraphics[scale=1]{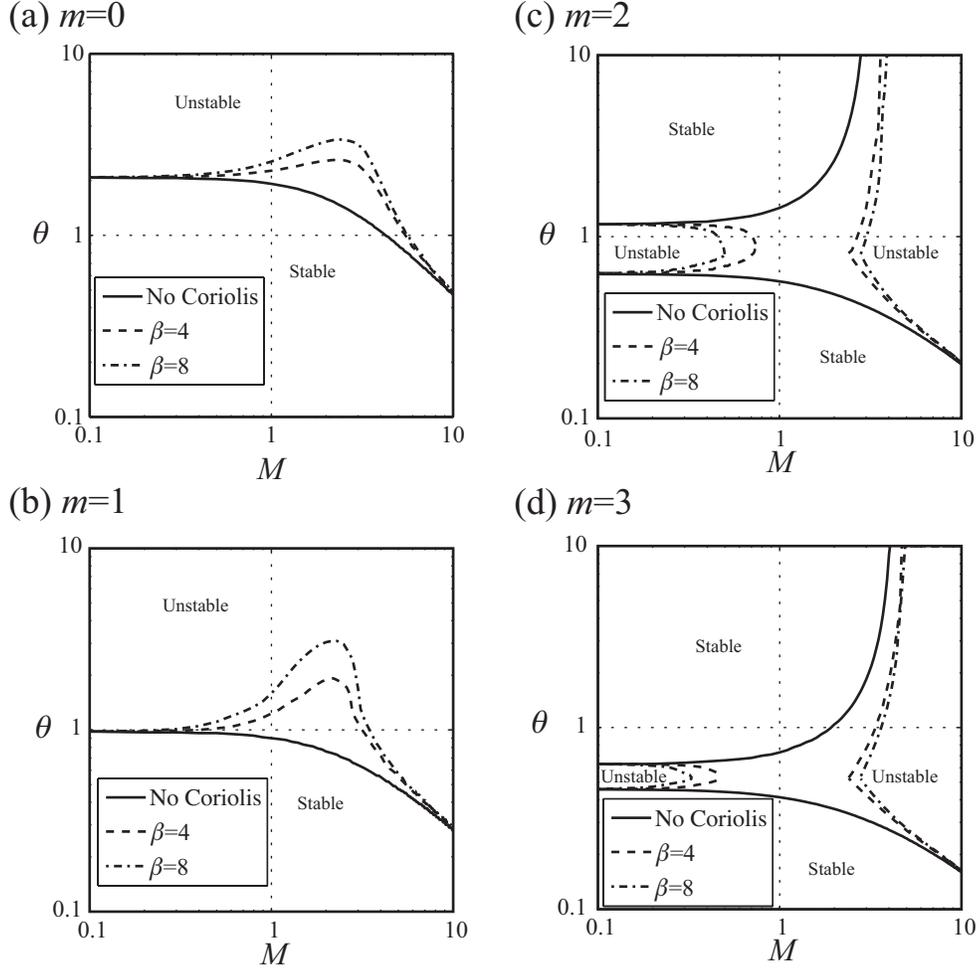}\\
  \caption{Marginal stability curves in cylindrical system calculated numerically from the eigenvalue problem given by Eqs. (\ref{cyl_xi}) and (\ref{cyl_BC}). Results for different azimuthal modes are shown:  (a) $m=0$, (b) $m=1$, (c) $m=2$, (d) $m=3$. Solid line denotes the case in which the Coriolis force is ignored; dashed line is for $\beta=4$; dash-dotted line is for $\beta=8$. In all cases the axial wave number is $k_z=-\pi/2$.}\label{cyl_m}
\end{center}
\end{figure}

To account for the Coriolis force we solve numerically the full eigenvalue problem given by Eqs. (\ref{cyl_xi}) and (\ref{cyl_BC}). The results for modes with different azimuthal mode numbers $m$ are presented in Figs. \ref{main}(II.a-c) and \ref{cyl_m}. It appears that the Coriolis force plays a significant stabilizing role in the range of Mach numbers $1<M<5$. This stabilizing effect is amplified even more with the increase of $\beta$ (Fig. \ref{cyl_m}). We also note the existence of unstable ``windows" on the plane of parameters $(M,~\theta)$ for modes with $m\geq2$. Such unstable ``windows" have been found in a similar screw pinch  configuration with rigid plasma rotation in Ref. \cite{Nijboer_1997}; however, the model used in that paper was not isothermal and equilibrium density was assumed constant. Our main finding is that the marginal stability curves for different $m$ follow the tension-mediated Parker instability scaling given by Eq. (\ref{Q_cr}) when Mach number $M>5$ [Fig.~\ref{main}(II.a)].  At these values of Mach number, the primary destabilizing mechanism in the system is  due to the Parker instability.

As one can see from Fig. \ref{main}(II.b), for non-axisymmetric modes with $m>0$ the growth rate of the instability becomes zero at some value of pinch parameter $\theta_0$ above the critical one,  $\theta_0>\theta_{cr}$. This value is determined by the condition $F=k_z+m\theta_0=0$, i.e., the corresponding mode is a pure interchange mode (perturbations do not bend the magnetic field lines). Such mode is a consequence of the assumed periodicity in $z$; it does not appear in the bounded cylinder (Sec. V), where characteristic wave number $k_z$ cannot be introduced. Fig. \ref{main}(II.c) shows that similar to the results in slab geometry, the critical pinch parameter $\theta_{cr}$ decreases with increasing azimuthal mode number $m$ when Mach numbers $M\geq5$.

\section{Bounded cylinder: dissipative MHD stability and nonlinear dynamics}

In this section we investigate the MHD stability of a rotating plasma screw pinch in the more realistic and experimentally relevant geometry of a bounded cylinder [Fig. \ref{geom}(b)], and assuming finite dissipation (here we take $\nu=\eta=0.01$). The equilibrium configuration is the same as in the ideal case in Sec. IV and given by Eqs. (\ref{cyl_eq}) and (\ref{cyl_n}). The stability study is performed numerically using the NIMROD code. 

Our first step in this stability study is to solve the eigenvalue problem corresponding to linearized Eqs. (\ref{mhd1})-(\ref{mhd3}) with boundary conditions given by Eqs. (\ref{bc_ideal}), (\ref{bc_nonideal}). Using NIMROD we solve the initial value problem and determine the growth rate of the fastest eigenmode. Since NIMROD uses the Fourier decomposition in periodic $\varphi$-direction \cite{Sovinec_2004}, and the Fourier harmonics are all decoupled in the linearized equations, we are able to determine the growth rates for all azimuthal mode numbers $m$. These results are presented in Fig. \ref{main}(III.a-c). 

Fig. \ref{main}(III.a) shows the calculated stability boundaries on the plane of Mach number--pinch parameter ($M-\theta$). The marginal stability curves for different $m$ follow the tension-mediated Parker instability scaling given by Eq. (\ref{Q_cr}) when Mach number $M>5$. In the region of small Mach numbers $M<1$,  the critical pinch $\theta_{cr}$ is practically independent of  $M$. Instabilities in this region of parameters are due to the curvature of the magnetic field lines and take the form of axisymmetric ``sausage" modes or non-axisymmetric kink modes. At medium values of Mach number $1<M<5$, the Coriolis force has a  stabilizing effect on both types of instabilities. The dependence of the Parker instability growth rate $\gamma$ on the pinch parameter $\theta$ for Mach number $M=6$ and different azimuthal numbers $m$ are presented in Fig. \ref{main}(III.b). According to these results, the characteristic time of the instability is the Alfvenic time scale $t_A=a/V_A$.
  
Fig. \ref{main}(III.c) shows the dependence of the critical pinch parameter $\theta_{cr}$ on the azimuthal mode number $m$ for several Mach numbers $M$, which are in the Parker instability region. We note that for every $M>5$ the critical pinch parameter $\theta_{cr}$ reaches a minimum for modes with  $m=4-5$.  These modes  determine the threshold of the Parker instability.  This is a consequence of two counteracting effects: the decrease of $\theta_{cr}$ with increasing $m$ when dissipation can be neglected [as seen in Figs. \ref{main}(I.c) and \ref{main}(II.c) for ideal cases], and the increase of $\theta_{cr}$ for modes with larger $m$ (shorter wave lengths) due to viscous and ohmic dissipation.   

\begin{figure}[tbp]
\begin{center}
  \includegraphics[scale=1]{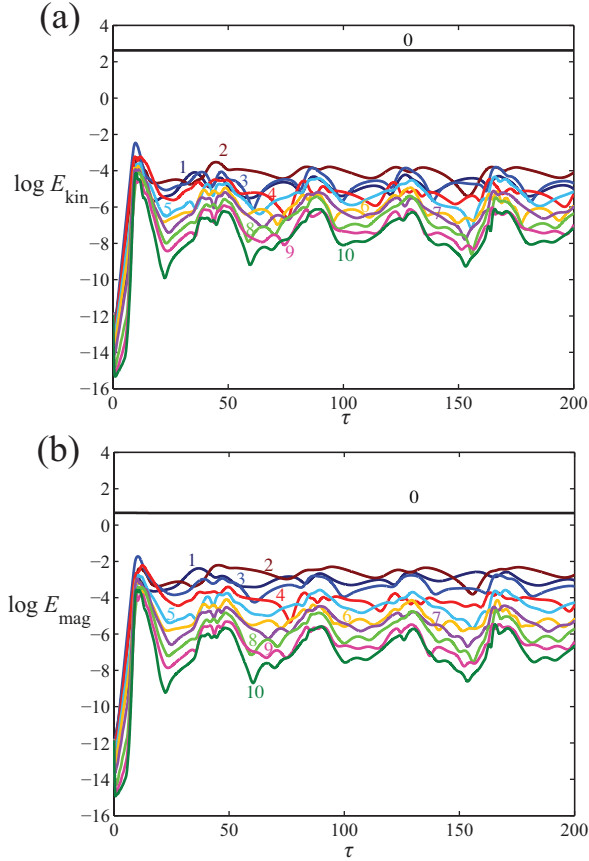}\\
  \caption{Time dynamics of (a) kinetic $E_\textrm{kin}$ and (b) magnetic $E_\textrm{mag}$ energies of different azimuthal modes (labeled) in nonlinear run with Mach number $M=6$, pinch parameter $\theta=1$, thermal to magnetic pressure ratio $\beta=8$, viscosity $\nu=0.01$ and resistivity $\eta=0.01$.}\label{energy}
\end{center}
\end{figure}

\begin{figure}[tbp]
\begin{center}
  \includegraphics[scale=1]{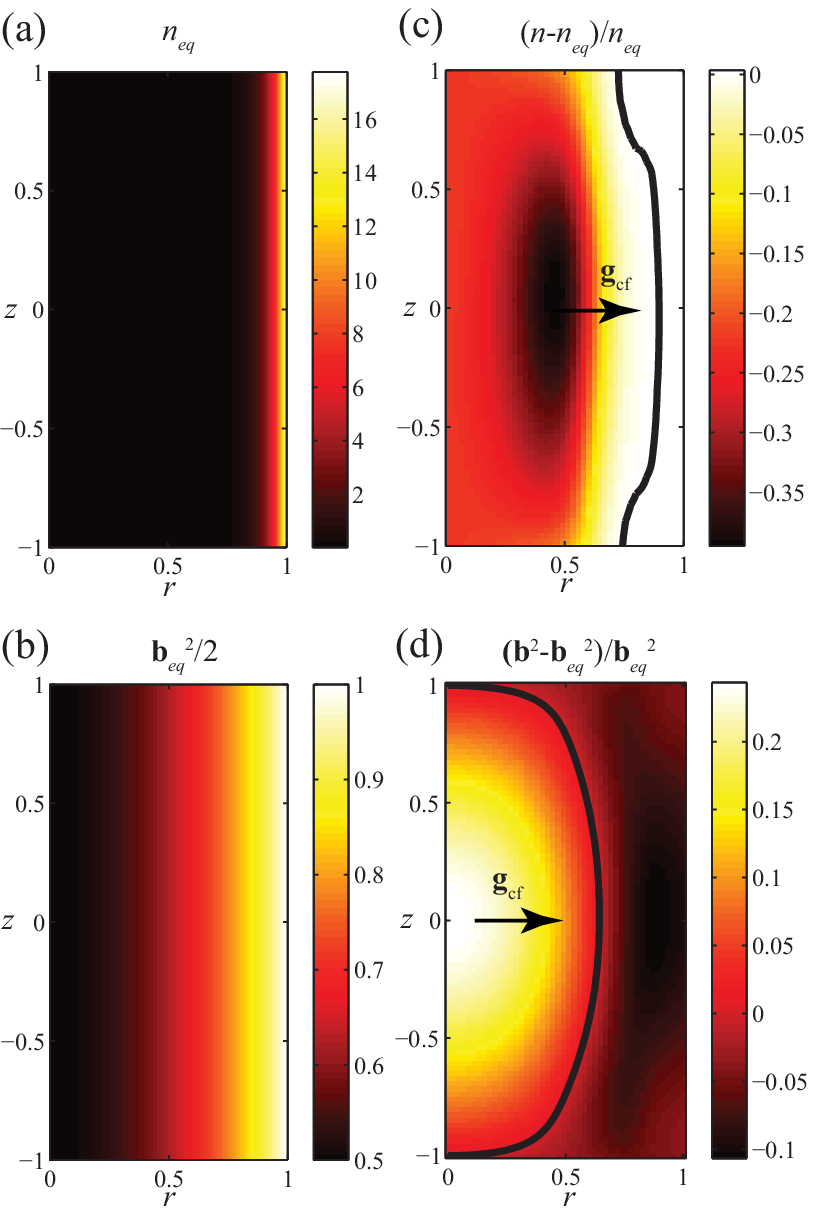}\\
  \caption{Relative deviations in $(r,~z)$ plane of the axisymmetric parts of (c) density and (d) magnetic pressure from their respective equilibrium profiles (a) and (b) during the nonlinear phase of the Parker instability. Solid black lines in (c) and (d) denote the points where deviation is zero. Arrows show the direction of the centrifugal acceleration $\g_\textrm{cf}$. Calculations are done with parameters listed in caption of Fig. \ref{energy}.}\label{final}
\end{center}
\end{figure}

Next, we report the results of 3-D simulations of the nonlinear development of the Parker instability. We solve the full system [Eqs. (\ref{mhd1})-(\ref{mhd3})] with the following parameters: $M=6$, $\theta=1$, $\beta=8$, $\nu=\eta=0.01$; at these parameters the system is unstable and the destabilization mechanism is dominated by the Parker instability.  Fig. \ref{energy} demonstrates the time dynamics of the kinetic and magnetic energies in the simulations. After the initial linear phase of the instability, the energies of the non-axisymmetric parts reach some average level. The back reaction of the non-axisymmetric distortions on the initial equilibrium configuration leads to its modification. Such modifications in axisymmetric parts of the density and magnetic pressure are shown in Fig. \ref{final}.  As expected, the nonlinear dynamics of the instability eliminates its original cause: plasma (``heavy" fluid) goes down along the effective centrifugal gravity, while the magnetic field  (``light" fluid) rises up against the gravity. This situation in many aspects is similar to the well known Rayleigh-Taylor instability.

\section{Conclusion}

We have performed the stability study of the analogue of the Parker instability (magnetic buoyancy) in a rigidly rotating plasma column with constant-pitch magnetic configuration.  The plasma rotation creates a centrifugal acceleration, which imitates the gravity required for the classical Parker instability. To distinguish this instability from instabilities related to magnetic field line curvature we have also considered the analogous plasma configuration in a slab geometry.

Applying the energy principle and solving the eigenvalue problem in cylindrical geometry, we found the marginal stability curves in the plane of Mach number--pitch parameter $(M-\theta)$. It appears  that the Parker instability  determines the stability thresholds for relatively large values of Mach number $M>5$; the scaling of the critical (required for instability) pinch parameter in this case is $\theta_{cr}\propto M^{-1}$ [Eq. (\ref{Q_cr})]. The mechanism of the Parker instability  has some analogy with the Rayleigh-Taylor instability: magnetic field (``light'' fluid) supporting plasma (``heavy" fluid) against the gravity is potentially buoyant and tends to rise up. 

At small values of Mach numbers $M<1$ the instability in the system is primarily due to the magnetic field line curvature (``sausage" for azimuthal number $m=0$ or kink instabilities for $m\ne0$). 

At medium values $1<M<5$ the Coriolis force stabilizes both types of instabilities; this effect depends on plasma $\beta$, having a stronger stabilization when $\beta$ is larger. The Coriolis stabilization of the kink-like instabilities in the systems with rigid-body rotation is somewhat known in the astrophysical community \cite{Carey_2009} but has not been actively studied in relation with experiments. At the same time, this stabilizing Coriolois effect could play an important role in modern experimental devices (e.g., tokamaks, reversed field pinches etc.)  where plasma rotation is commonly observed.    

As follows from our ideal MHD analysis, for a given Mach number $M>5$ the onset of the Parker instability depends only on the value of normalized gradient of the equilibrium magnetic field (pinch parameter $\theta$) but not on its magnitude. This means that the instability can be obtained even for infinitely small magnetic field.  However, in this case the growth rate of the instability approaches zero, because the growth rate scales with the minimized potential energy [Eq.~(\ref{cyl_Wmin})] which itself scales as $B_{z0}^2$ and thus $\omega^2\propto W_{min}\propto B_{z0}^2$. Therefore, to observe the Parker instability in experiment a relatively small magnitude of the equilibrium magnetic  field is sufficient; in fact, it is just enough to have a field exceeding the Earth's magnetic field. Our analysis has been conducted for isothermal motions in an isothermal plasma, as should be appropriate for conditions realized in MPCX.  If thermal conductivity is decreased, this analysis likely provides a lower bound on the Parker instability.

Our results suggest that the Parker instability can be achieved in an experiment with controllable plasma flows, such as MPCX. Simple estimates show that in order to reach the Mach number of $M=6$, one may have an argon plasma with $\mu_i=40$, electron temperature  $T_e=4$ eV and peak velocity of $V_0=20$ km/s. With an applied axial magnetic field of $B_0=4$ Gauss, the total axial current required for instability is $I_\textrm{total}=1$ kA; for a uniform distribution of current this will give a pinch parameter of $\theta=1$. These parameters may soon be achievable  in the MPCX.      

\acknowledgements

The work is supported by the Center for Magnetic Self-Organization in Laboratory and Astrophysical Plasmas. Benjamin Brown is supported in part by NSF Astronomy and Astrophysics postdoctoral fellowship AST 09-02004.  Noam Katz acknowledges support from the Department of Energy and the Center for Momentum Transport and Flow Organization.

\end{document}